\documentstyle[prl,twocolumn,aps]{revtex}

\begin{document}
\draft

\title{Teleportation via generalized measurements,
and conclusive teleportation}

\author{
Tal Mor\thanks{Electrical Engineering department, University of California,
Los Angeles, CA 90095, USA}
and
Pawel Horodecki\thanks{Faculty of Applied Physics and Mathematics,
Technical University of Gda\'nsk, 80--952 Gda\'nsk, Poland}
}

\date{\today}

\maketitle

\begin{abstract}

In this work we show that teleportation~\cite{BBCJPW}
is a {\em special case} 
of a generalized 
Einstein, Podolsky, Rosen (EPR) non-locality.
Based on the connection between teleportation and generalized measurements
we define conclusive teleportation.
We show that perfect conclusive teleportation 
can be obtained with any pure entangled state, and 
it can be arbitrarily approached
with a particular mixed state.

\end{abstract}

\noindent
KEY WORDS: Quantum information processing, Entanglement, Teleportation,
Nonlocality, Generalized measurements, Conclusive teleportation,
Distillation;

\section{Introduction}

Quantum information processing 
(QIP)~\cite{Hels67,Kholevo1,Kholevo2,Helstrom,Davies,Per93,Ben-Div-95,Ben-Sho-98}
discusses information processing in which 
the basic units are two-level quantum systems (e.g., spin-half particles,
the polarization of individual photons, etc.) 
known as quantum bits or shortly,
{\em qubits}.
The classical states, $0$ and $1$, of a classical bit 
are generalized to quantum states
of a qubit, $|0\rangle \equiv { 1 \choose 0}$ and 
$|1\rangle \equiv { 0 \choose 1}$.
The nonclassical aspect of a qubit is that it can also be in a
superposition
$|\phi \rangle = \alpha |0\rangle + \beta |1\rangle = { \alpha \choose \beta}$,
with $|\alpha^2| + |\beta^2| = 1$, and two or more qubits can be in a
superposition which cannot be written as tensor product, and is  known as
an entangled state. The special properties of entangled states were first noted
by Einstein-Podolsky-Rosen (EPR)~\cite{EPR}, and a proof for the
special nonclassicality was first obtained by Bell~\cite{Bell}.
The EPR-Bohm singlet state, 
$ |\Psi^-\rangle = (1/\sqrt2) [|01\rangle - |10\rangle]$,
of pair of qubits is the
most important example of entanglement. [We prefer,
for simplicity, to use these ``braket'' notations for two-particle states
while using vector notations for one-particle states.]
The singlet state can be complimented to a
basis~\cite{BMR} (known now as the ``Bell basis'') by adding the three states
$ |\Psi^+\rangle = (1/\sqrt2) [|01\rangle + |10\rangle]$,
$ |\Phi^-\rangle = (1/\sqrt2) [|00\rangle - |11\rangle]$,
$ |\Phi^+\rangle = (1/\sqrt2) [|00\rangle + |11\rangle]$,
the Bell-states (or the Braunstein, Mann, Revzen (BMR) states).
We shall usually refer here to two qubits in any
one of the Bell-BMR states as an
EPR-pair, and to the EPR-Bohm state as the singlet state. 
Entanglement---the quantum feature visualized
by such states---is an origin of
fascinating quantum phenomena in quantum information theory:
quantum computation \cite{deutsch85,ShorFactor,Grover,Ben-Div-95,NatureL},  
entanglement-based quantum cryptography \cite{EPR-scheme,BHM96},
quantum error
correction~\cite{ShorQEC,ScienceCLSZ,ShorFT,ScienceZ,CRSS,Rains}, and more. 

One of the most fascinating discoveries
is {\it quantum teleportation}~\cite{BBCJPW} 
which lies in the heart of quantum information theory 
(see~\cite{Ben-Sho-98,Caves}),
and has been recently realized
experimentally
\cite{ExTel}.  Quantum teleportation is
a process of transmission of an {\it unknown}
quantum state $|\phi \rangle={\alpha \choose \beta}$
via a previously shared EPR pair with the help of only two classical bits
transmitted via a classical channel (usually visualized by phone):
Alice
(the sender) has a qubit 
in an unknown quantum state 
which she wishes to transmit to Bob (the receiver)
using additional EPR pair shared by her and Bob.
To do this she performs joint measurement on the two particles
which are in her hands,
then she sends (via phone) her two-bit result to Bob, who
performs some unitary operation on his particle
``transforming''  his particle to the (still unknown) original state
$|\phi\rangle$.
The initial state of Alice's unknown state, and the EPR-pair
(say, in a singlet state) is 
$ {\alpha \choose \beta} | \Psi^- \rangle $. 
The teleportation is based on the fact that this initial state can also be
written as~\cite{BBCJPW}:
\begin{eqnarray}
&&|\Psi_{123}\rangle  = 
{\alpha \choose \beta}_1 | \Psi^-_{23} \rangle \ =
\frac{1}{2} \Big[ | \Phi^+_{12} \rangle { -\beta \choose  \alpha}_3 +
 | \Phi^-_{12} \rangle { \beta \choose \alpha}_3 + \nonumber \\
&& | \Psi^+_{12} \rangle { - \alpha  \choose  \beta }_3 +
 | \Psi^-_{12} \rangle { - \alpha  \choose -  \beta }_3 \Big]  \ ,
 \end{eqnarray}
where we add the particle's numbers to avoid confusion.
A Bell measurement at Alice's site projects the state of Bob's particle,
to be in one of the states 
\begin{equation}
{ -\beta \choose  \alpha}\ ; \ 
{ \beta \choose \alpha}\ ; \ 
{ - \alpha  \choose  \beta }\ ; \ 
{  \alpha  \choose  \beta }  \ ,  
\end{equation}
depending on Alice's outcome.
Using the appropriate rotation,
\begin{eqnarray}  
 \left( \begin{array}{cc}
                          0   & 1 \\ -1 & 0  
\end{array} \right)  \ ;
 \left( \begin{array}{cc}
                          0   & 1 \\ 1 & 0  
\end{array} \right)  \ ;
 \left( \begin{array}{cc}
                          -1   & 0 \\ 0 & 1  
\end{array} \right)  \ ; or
 \left( \begin{array}{cc}
                          1   & 0 \\ 0 & 1  
\end{array} \right)  \ , 
\end{eqnarray}
respectively,
each of these states can be rotated back to yield the 
unknown state $|\phi\rangle$. Bob chooses the correct rotation based on the two
bits he receives from Alice.

The minimal resources required for teleportation are one
EPR singlet pair, which clearly, is 
{\em independent} of $|\phi\rangle$,
and two classical bits. This seems to be rather mysterious
because
(i) the particle is described by a point on a unit
sphere, hence by two {\it real} numbers and not by two bits,
(ii) as one can check, even from those two classical bits
neither Alice nor Bob can learn anything about the unknown parameters
of the state $|\phi\rangle$.

The alternative approach presented in this paper somewhat
clarifies the mystery. 
Namely, we interpret the teleportation in the
light of the paper 
of Hughston, Jozsa and
Wootters (HJW)~\cite{HJW}, 
and we present the teleportation process as a unique case of 
generalized EPR-nonlocality (we use the language of generalized measurement
to express the ideas of~\cite{HJW}).

A positive operator valued measure (POVM) provides the most general 
physically realizable measurement in quantum 
mechanics~\cite{JP-DL,Helstrom,Davies,Per93}, 
and we also call these measurements
``generalized measurements''. Formally, a POVM is a collection of
positive operators $A_i$ on a Hilbert space ${\cal H}_n$ of dimension $n$
which sum up to the identity, $A_1 + \ldots + A_r = I_n$.
[When viewed as matrices, these are matrices which can be
diagonalized and have only non-negative eigenvalues.]
Standard measurements (which are usually described by some Hermitian
operator in quantum mechanics books) arise as a special case where
$A_i = |\psi_i\rangle \langle\psi_i|$ and $A_i A_j = \delta_{ij}$.
We discuss here only pure POVMs in which each of the $A_i$ is 
proportional to a projection
$A_i = q_i |\psi_i\rangle \langle\psi_i|$, but the operators $A_i$
are not necessarily orthogonal to each
other, so that $r \ge n$.
Any POVM can be implemented (at least in principle) by adding an ancilla in
a known state, and performing a standard measurement in the enlarged
Hilbert space~\cite{Per93}.

To describe the EPR-nonlocality and its generalization, let us first define
the notion of $\rho$-ensembles~\cite{HJW}.
An ensemble of quantum state is defined by a collection of normalized
states $|\psi_1\rangle, \ldots,  |\psi_m\rangle$ taken with a-priori
probabilities $p_1, \ldots, p_m$ respectively.
To any such ensemble one can associate its density matrix:
\begin{equation} 
\rho = \sum_{i=1}^m p_i |\psi_i\rangle \langle \psi_i|
\ , 
\end{equation}
and the term $\rho$-ensemble refers to an ensemble with a density matrix
$\rho$.
For instance, for the completely mixed state in 2-dimensions,
$\rho = I/2$, the following are all legitimate $\frac{I}{2}$-ensembles:
\begin{eqnarray}
E_1 &=& \{|\psi_1\rangle = {1 \choose 0}, 
|\psi_2\rangle = {0 \choose 1}; p_1 = p_2 = 1/2 \} \nonumber \\
E_2 &=& \{|\psi_1\rangle = {1/\sqrt2 \choose 1/\sqrt2}, 
|\psi_2\rangle = {1/\sqrt2 \choose -1/\sqrt2}; 
p_1 =  p_2 = 1/2 \} \nonumber \\
 E_3 &=& \{|\psi_1\rangle = {1 \choose 0}, 
|\psi_2\rangle = {0 \choose 1}, 
|\psi_3\rangle = {1/\sqrt2 \choose 1/\sqrt2}, 
\nonumber \\
&&  
|\psi_4\rangle = {1/\sqrt2 \choose -1/\sqrt2}; \
p_i = 1/4 \ ,  \ 
1 \leq i \leq 4
\} \nonumber \\
 E_4 &=& \{
|\psi_1\rangle = {\alpha \choose \beta  }, 
|\psi_2\rangle = {\alpha \choose -\beta  },  
|\psi_3\rangle = {\beta  \choose \alpha }, 
\nonumber \\
&&  
|\psi_4\rangle = {\beta  \choose -\alpha }; \
p_i = 1/4 \ , \ 
1 \leq i \leq 4
 \}\ .
\end{eqnarray}

When a classical system is subjected to a measurement of any of its
properties a definite outcome exists (at least in principle).
However, 
when a quantum particle (say a qubit) 
is in a state which is well defined in one
bases, say ${1 \choose 0}$ in the rectilinear basis
${1 \choose 0}; {0 \choose 1}$,
the state is undefined in any other basis,
and a measurement, say, in the diagonal basis
${1/\sqrt2 \choose \pm 1/\sqrt2}$, does not have
a definite outcome which can be predicted, 
and only the probabilities (of the possible outcomes) can be
calculated. This is the well known {\em uncertainty principle}.

The EPR paradox~\cite{EPR} is as follows: 
If Alice and Bob share a singlet state,
the state of Bob's particle is undefined (if we trace-out Alice's particle,
then Bob's particle is in a completely mixed state $I/2$, but without
tracing out Alice's particle, the state of Bob's particle by itself is not
defined).
However, if Alice measures in any basis she chooses to, say 
the rectilinear or the diagonal,  
she fully ``learns''
the state of Bob's
particle. Assuming that a quantum state is ``real'' (as the state of a
classical object) and assuming that the state cannot be changed 
instantaneously (immediately after Alice's measurement)
when Alice and Bob are far apart, EPR concluded that 
the state of Bob's particle must have been previously defined in 
{\em both}
bases, in contradiction with the uncertainty principle.
They further concluded that 
this is a paradox (the EPR paradox) 
and thus that quantum mechanics is incomplete.
Today we know, due to~\cite{Bell},
that indeed quantum mechanics
is not described by a realistic-local model, and thus the EPR-paradox is
resolved.

We refer to the following fact as the EPR nonlocality:
the state of Bob's particle, previously
undefined, become completely specified by Alice nonlocal operation. 
Thus, the EPR nonlocality is not a nonlocality in the sense of~\cite{Bell},
but the profound feature 
which allows to ``create'' quantum states
from different ensembles as it was discussed in the original EPR analysis.
  
Using the language of $\rho$-ensembles, the EPR nonlocality is described as
follows: 
Alice can choose whether Bob's
state will be in a $\rho$-ensemble $E_1$ or $E_2$ by choosing an appropriate
measurement on her member of the EPR pair.
Thus, while Bob holds the mixed state $\rho$, Alice has an additional
information regarding his state. 

The EPR nonlocality is further generalized by 
HJW in~\cite{HJW}, by allowing Alice to perform generalized
measurements (POVMs), hence enabling her to create {\em any}
$\rho$-ensemble in Bob's site, and also knowing precisely the state he has.
Note that she cannot chose $\rho$, and also she cannot chose 
the resulting state in Bob's hands, but she can choose the $\rho$ ensemble,
and learn the state.
Generating $\rho$-ensembles at a distance is the generalization 
of the EPR nonlocality in which only {\em standard (projection)
measurements} are used. We shall refer to this generalized EPR nonlocality
as the EPR-HJW nonlocality.

In particular Alice can create the $\rho$-ensemble $E_4$, and we shall show
in Section~III,
that creating this ensemble corresponds to the teleportation process,
once we add the transmission of classical information from Alice to Bob
(she transmits the outcome of her measurement).
Thus, teleportation
is a special case of generating
$\rho$-ensembles at a distance, when Alice
uses a special POVM and where the operations done by Alice and Bob
are independent of the parameters of the (unknown) state.
We call this view of the teleportation process ``telePOVM''
(see the acknowledgement), or
teleportation via generalized measurements.

The next natural step is to use this approach to generalize
the concept of teleportation, by removing the demand that
the transmitted state can always be recovered.
In Section~IV, we define the concept of {\em conclusive teleportation}.
The term ``conclusive'' is taken from quantum information theory, when
one asks the following question (see, for instance, \cite{Per93}):
what is the optimal mutual information which can be extracted from
two nonorthogonal quantum states each sent with probability half?
One can obtain a definite (correct) answer (regarding
the given state) {\em sometimes}
for the price of knowing nothing 
in other occasions~\cite{Per93}.
Here we adapt this term presenting the conclusive teleportation
in which the teleportation process is {\em sometimes} successful, and the
sender knows if it is successful or not.
When Alice and Bob use an 
entangled pure state which is not fully entangled
the conclusive teleportation scheme allows them to teleport a quantum state
with fidelity one.
This is done for the price of occasional failures, and
the sender knows whether it is successful or it is not.
For many 
purposes (e.g., for quantum cryptography~\cite{BB84,EPR-scheme,Ben92,BHM96}),
one would prefer performing this conclusive teleportation rather than
the original
one which leads to a transfer fidelity which is smaller than 
one~\cite{Gisin}, and yields fidelity one only when 
the shared state is maximally entangled. [The fidelity of a state
$\rho$ relative to a pure state $|\psi\rangle$ is given by
$\langle \psi|\rho|\psi\rangle$; for other properties of the
fidelity, see~\cite{Fuc-Gra-99}.]
For instance, if Alice has an unknown qubit which she wishes to 
teleport to Bob, while they only 
share partially-entangled states, she can first create a fully entangled
state, and try to teleport a qubit from such a pair to Bob via a conclusive
teleportation. If she fails, she can try again (using another shared pair)
till she succeeds.
Once she succeeds to teleport an EPR-pair member, 
she can teleport the unknown qubit
with fidelity one.

A further generalization is to let Bob also perform a conclusive measurement
that sometimes succeeds (this requires, a 2-way classical communication).
Surprisingly, we shall show in Section~V, that
this type of teleportation can allow for a conclusive
teleportation even when the shared entangled state is {\em mixed}.
The conclusive teleportation obtained in this case is with arbitrarily
high fidelity, but for the price of a probability of success
decreasing to zero as the fidelity increases.
We refer to it as a quasi-conclusive
teleportation. The questions of quasi-conclusive teleportation
with fixed probability of success or with only one-way
classical communication allowed will be discussed elsewhere.

\section{TelePOVM}

Suppose that Alice and Bob share
any two-particle entangled 
pure state in any dimension, such that the reduced density
matrix in Bob's hands is $\rho$.
Then, according to Hughston, Jozsa and Wootters~\cite{HJW}, 
any measurement at Alice side, performed on her part of the
entangled state, creates a specific $\rho$-ensemble in Bob's hands.
All $\rho$-ensembles are indistinguishable (recall that a quantum system
is fully described by its density matrices)
unless there exist an additional information somewhere.
For example, in the Bennett-Brassard-84 (BB84)
cryptographic scheme~\cite{BB84}
Bob receives the same density
matrix $\rho$ whether Alice uses the rectilinear basis or the 
diagonal basis, but he receives
different $\rho$-ensembles. He cannot distinguish between the two ensembles
and between the states in each particular occasion,
unless he receives more information from Alice.
When receiving additional information (the basis) he is told which
$\rho$-ensemble he has, and (in this particular case)
can find which state.

In
the same sense, the EPR-scheme\cite{EPR-scheme}, provides a simple example
of the HJW meaning of $\rho$-ensembles:
when Alice chooses to measure her member of the singlet
state in the rectilinear basis or in the diagonal 
basis, she ``creates'' a different
$\rho$-ensemble in Bob's hands, $E_1$ or $E_2$ respectively.
Bob can distinguish the two states
to find Alice's bit
after receiving additional information from Alice who tells him the basis
(hence her choice of a $\rho$-ensemble).
Alice's choice of measurement determines the
$\rho$-ensemble, and furthermore, her result in each occasion, tells
her which of the states is in Bob's hands.
If the measurement is chosen in advance, and Alice tells Bob the outcome 
of the measurement (by sending one bit of information) he can know precisely
the state of the qubit in his hands.

The generalization done by HJW
replaces the standard,
projection measurement by a generalized
measurement~\cite{JP-DL,Helstrom,Per93} (POVM),
so the number of results can be larger than the dimension
of the Hilbert space in Alice's site or in Bob's site.
Thus, the HJW-EPR nonlocality argument
implies that the set of Bob states contains
nonorthogonal states. Furthermore, if
Alice sends him an additional information
(her measurement's result) Bob can recognize in which of these states 
his particle is now.
This is a very interesting result of~\cite{HJW}
and we now show that teleportation provides a fascinating usage of it.

Let Alice and Bob share an EPR pair (say, the singlet state).
Consider the following POVM ${\cal A}$:
\begin{eqnarray}  
 A_1 &=& \frac{1}{2}  \left( \begin{array}{cc}
                          \alpha^2 & \beta\alpha^* \\ \beta^*\alpha & \beta^2
\end{array} \right)  \ ;
  \    A_2 = \frac{1}{2}  \left( \begin{array}{cc}
                          \beta^2 & -\beta^*\alpha \\ -\beta\alpha^* & \alpha^2
\end{array} \right)  \ ; \nonumber \\
 A_3 &=& \frac{1}{2}  \left( \begin{array}{cc}
                          \beta^2 & \beta^*\alpha \\ \beta\alpha^* & \alpha^2
\end{array} \right)  \ ;
  \    A_4 = \frac{1}{2}  \left( \begin{array}{cc}
                         \alpha^2 & -\beta\alpha^* \\ -\beta^*\alpha & \beta^2
\end{array} \right)  \ ,
\label{POVM}
\end{eqnarray}
with complex
parameters $\alpha$, $\beta$, such that $|\alpha|^{2}
+|\beta|^{2}=1$.
These matrices
have positive eigenvalues and
sum up to the unit matrix therefore form a POVM.
Following the arguments of HJW,
applying such a POVM to one member of two particles
in an EPR state is equivalent to a choice of a specific
$\rho$-ensemble combined of four possible states;
when the result of the POVM is $A_i$, the other member
is projected onto a state orthogonal to $A_i$,
i.e., it will be in one of the states
$\psi_1 = {\beta \choose -\alpha}\ $;
$\psi_2 = {\alpha \choose \beta}\ $;
$\psi_3 = {\alpha \choose -\beta}\ $, and 
$\psi_4 = {\beta \choose \alpha}$ respectively, and Alice will know in which
of them. Alice can send Bob two-bit information to describe the outcome of
her measurement (one of her four results),
and this information actually tells
him which of those four states he got.
Then Bob can re-derive one of the states, say ${\alpha \choose \beta}$,
by performing the appropriate rotation,
according to the two classical bits he is being told.
The reason that exactly two bits are required here is that the POVM has
four outcomes.

It should be stressed that, for this specific POVM (\ref{POVM}),
Bob's recovering operations {\it do not} depend on the
parameters $\alpha$, $\beta$, so these need not be known to him. 

Every POVM can be performed in the lab by performing a standard measurement
on the system $\rho_{sys}$, plus an ancilla \cite{IvPe,Per93}
(this is a property of the POVM so it is true independently of
the state of the measured system).
One way to perform the POVM (\ref{POVM}) is to take an ancilla in a 
state $\phi={\alpha \choose \beta}$, and perform the Bell measurement
(a measurement such that the outcomes are the Bell-BMR states)
on the ancilla and the system.
The first operator, $A_1$,
results from the measurement of the projection operator
$P_1 =  | \Phi^+ \rangle \langle \Phi^+ |$ in the Hilbert-space 
of Alice's particle
plus the ancilla.
Applying the technique described in~\cite{Per93} (Chapter 9,
sect. 9.5, 9.6, about generalized measurements
and Neumark's theorem)
we get terms of $A_{1}$
\begin{equation} (A_1)_{mn} = \sum_{rs} (P_1)_{mr,ns} (\rho_{aux})_{sr}
\ , \end{equation}
where $\rho_{aux}$ is the state of the ancilla, the $mn$ are the
indices of the particle
and the $sr$ are indices of the ancilla.
The $m=0,\ n=0$ case corresponds to multiplying the
upper left block of $P_1$ by the density matrix of the ancilla, and
tracing the obtained matrix yielding:
\begin{equation} {\rm Tr} \  
\left(\begin{array}{cc} \frac{1}{2} & 0 \\ 0 & 0
\end{array}\right)_{rs}
\left(\begin{array}{cc} \alpha^2 & \beta^*\alpha \\
\beta\alpha^* & \beta^2
\end{array}\right)_{sr} = \frac{1}{2} \beta^2 \ .
\end{equation}
The $m=1,\ n=0$ case (second line, first column, in $A_1$) results from a
similar multiplication but with the lower left block of $P_1$.
In the same way we calculated the other elements of that
operator, and the other three operators, and we verified that the
Bell measurement corresponds to the desired POVM.

It should be stressed that 
Alice's measurements {\em do not} depend on the parameters
$\alpha$, $\beta$, thus these need not be known to her.
Moreover she can learn nothing about the latter as all four results
of her generalized measurement corresponding to operations (\ref{POVM})
can happen with equal probabilities.
In the case of starting with the singlet state,
all four Alice's results occur with
equal probabilities and the initial state of Bob's
particle is the maximally mixed
state $\frac{I}{2}$
(reduced state of a maximally entangled state).
Thus, it is clear that the teleportation is equivalent 
to the creation of a specific
$\rho=\frac{I}{2}$ ensemble 
at a distance, where the specific $\frac{I}{2}$-ensemble is 
$E_4$.
This can be done even if Alice and Bob do not know the state of the ancilla,
${\alpha \choose \beta}$ chosen by someone else, 
and this is exactly the process of teleportation
of an unknown state.

This process
will also teleport a density
matrix (a mixed state) or a particle entangled with others.
It can also easily be generalized to fully
entangled states in higher ($N^2$) dimensions discussed in \cite{BBCJPW}.

\section{Generating $\rho$-ensembles in quantum key distribution}

To see one application of the ideas described above,
let us view a different
scenario (taken from quantum key distribution):
Suppose that Alice has in mind a set
of states and their probabilities, 
say, $E_3$, which is used in the BB84~\cite{BB84}
quantum key distribution scheme. This describes a particular $\rho$-ensemble
(the $\frac{I}{2}$-ensemble in the BB84 case) sent to Bob.
If Alice doesn't care which of the states is sent in each experiment, but only
that it belongs to that set, she does not
need to send the states. Instead of sending Bob the states, she sends him
a member of some entangled state such that the reduced density matrix in Bob's
hands is $\rho$.
Then she applies a specific POVM which creates the desired ensemble in Bob's
hands.
The relevant example is the EPR scheme~\cite{EPR-scheme}, in which an
EPR-pair is shared by Alice and Bob.
As we have seen before, Alice creates either the $\frac{I}{2}$-ensemble
$E_1$ or $E_2$, when she apply a measurement in the rectilinear or the
diagonal bases respectively.
However, since the probability of each basis is 1/2, Alice's 
full operation, including the choice of the basis,
can also be described by a POVM which leads to the ensemble $E_3$.

Let us present a less trivial example.  Let the state
\begin{equation}
| \chi_{23} \rangle = a | 00 \rangle +
              b | 11 \rangle \
\label{entstate}
\end{equation}
(with $a$, $b$ real, and $a^2+b^2=1$)
be prepared by Alice and let one particle be sent from Alice to Bob.
Then let Alice measure her particle 
using a standard measurement in the computation
(the rectilinear) basis. As result,
the following $\rho$-ensemble is generated in Bob's hands:
$\{  \frac{1}{\sqrt 2} {a+b \choose a-b},
\frac{1}{\sqrt 2} {a-b \choose a+b} ; p_1 = p_2 = 1/2 \}$.
This operation produces the Bennett-92 \cite{Ben92} 
scheme for quantum key distribution,
in the same way that the EPR scheme produces the BB84 scheme.

\section{Conclusive teleportation with any pure entangled state}

We first present the use of an additional 
one-way classical communication to modify  
the teleportation process: if Alice wishes to teleport to Bob 
a quantum state of which she can make more copies (e.g., to teleport a
member of an EPR-pair) or if she wishes to teleport
an arbitrary state from a set (e.g., a BB84 state), she can improve the 
teleportation process very much 
by using conclusive teleportation: a teleportation
process which is sometimes successful. 
After performing her measurement, Alice uses the classical
channel to tell Bob if the teleportation succeeded, and he uses the
received state only if the teleportation succeed.

For instance,
one can use conclusive teleportation to save time or classical bits.
Let Alice and Bob share a fully entangled state, and use it
to perform a conclusive teleportation:
Alice performs a measurement which distinguishes the singlet state from
the other three (triplet) states. Instead of sending $2$ bits she sends Bob
only one bit telling him whether she received a singlet state or not.
Bob doesn't need to do any operation on his particle.
In a $\frac{1}{4} $ of the occasions she receives this result (the singlet
state),
hence performs a successful teleportation.
This process makes sense when the classical bits are as expansive as the
shared quantum states, or when a fast teleportation of arbitrary states
(e.g. BB84 states) is required.
Also, it allows 
teleportation when Bob is technologically limited and 
cannot perform the required rotations.

The process of conclusive teleportation makes more 
sense when Alice and Bob share a pure entangled
state which is not fully entangled.

Let Alice and Bob share the state
(\ref{entstate})
({\it any} pure state can be written in that form 
called the Schmidt decomposition~\cite{HJW,Per93}),
which they use to teleport a quantum state 
$\phi_{1}={\alpha \choose \beta}_1$.
Following the method of \cite{BBCJPW}, the state of the three particles
is written
using the Bell-BMR
states as:
\begin{eqnarray}
&& | \Psi_{123} \rangle =  | \phi_1 \rangle  | \chi_{23} \rangle \ =
\frac{1}{\sqrt2} \Big[ |  \Phi^+_{12} \rangle {a \alpha  \choose b \beta }_3 +
 |  \Phi^-_{12} \rangle { a \alpha  \choose - b \beta }_3 + \nonumber 
\\
&&\ \ \ \  |  \Psi^+_{12} \rangle {a \beta \choose b \alpha}_3 +
 |  \Psi^-_{12} \rangle { -a \beta \choose b \alpha}_3 \Big] \ .
\end{eqnarray}
If Alice and Bob were to use the standard teleportation process, 
a Bell measurement still creates the same POVM as before.
But, unlike the case of using a fully entangled state,
the states created in Bob's hands depend also on $a$ and $b$,
and not only on the state of the ancilla.
The fidelity is clearly less than one (e.g., if Alice received a
state $\Phi^+$ in her measurement (which happen with probability
$p_{\Phi^+} = 
(|\alpha|^2 a^2 + |\beta|^2 b^2)/2$, 
the fidelity
$|\langle \phi_1 | \phi_1^{\rm out}\rangle|^2$ 
of the output state is
$ (|\alpha|^2 a + |\beta|^2 b)^2/ 
(|\alpha|^2 a^2 + |\beta|^2 b^2)$, which depends on $a$ and $b$, and 
on the teleported state. 

The POVM that reproduce the four desired states
can be found. It is not performed by a Bell measurement and
will depend on the state of the ancilla
which is supposed to be unknown to both sides.
So perfect teleportation will not take place this time.

We present a different measurement which generates the desired states in Bob's
hands with perfect fidelity.
The price we pay for the perfect state obtained, is that the process
cannot be done with 100\% probability of success, therefore it is a conclusive
teleportation.
To explain how it works,
let us return to the case of fully entangled state (standard teleportation,
with initial EPR-pair $\Phi^+$) and
separate the Bell measurement into two measurements (one follows the other):
\begin{enumerate}
\item
A measurement which checks whether the state is in the subspace
spanned by $ | 00\rangle $ and
$ | 11 \rangle $, or in the subspace spanned by
$ | 01 \rangle $ and
$ | 10 \rangle $.
\item
A measurement in the appropriate subspace (according to the result of the
previous step), which projects the state on one of the two possible
Bell states in that subspace, $\Phi^\pm$ and $\Psi^\pm$ respectively.
\end{enumerate}
When $| \Psi_{23} \rangle$
is not fully entangled we still repeat
the first step of that two-steps process.
To see the outcome, note that the
state of the three particles can also be written as
\begin{eqnarray}
&&| \Psi_{123} \rangle  =
    \frac{1}{2} \large[ [a | 00  \rangle
            + b  | 11 \rangle ] {\alpha \choose \beta}_3
+    [a | 00  \rangle
            - b  | 11 \rangle ] {\alpha \choose -\beta}_3
\nonumber \\
&&+    [b | 01 \rangle
            + a  | 10  \rangle ] {\beta \choose \alpha}_3
+    [b | 01 \rangle
            - a  | 10   \rangle ] {-\beta \choose \alpha}_3  \large]
 \ .\end{eqnarray}
The first step projects $|\Psi_{123}\rangle$
on either the first two possibilities or
the last two with equal probabilities.
In the second step, let us assume that the result of the first step was
the subspace spanned by the states 
$|00 \rangle \equiv {1 \choose 0}_{\{00;11\}}$
and $| 11 \rangle \equiv {0 \choose 1}_{\{00;11\}}$.
[A similar analysis can easily be done for the other case where the result of
the first step is
the subspace spanned by the states 
$| 01 \rangle \equiv {1 \choose 0}_{\{01;10\}}$
and $| 01 \rangle \equiv {0 \choose 1}_{\{01;10\}}$.]

In this 
${\{00;11\}}$
subspace, Alice now performs a second measurement, but not in
the Bell-BMR basis which is now the states 
$(1/\sqrt2){ 1 \choose \pm 1}_{\{00;11\}}$,
as in the ideal case. 
Instead, Alice performs
a POVM which conclusively distinguish the two states,
${a \choose b}_{\{00;11\}} $ and $ {a \choose - b}_{\{00,11\}}$
(which are the first two states in the above expression).
Assuming (without loss of generality) that $a^2 \ge b^2$
the POVM elements in that subspace are:
\begin{eqnarray}  A_1 &=&  \left( \begin{array}{cc}
                          b^2 & ba \\ ba & a^2
\end{array} \right)  \ ;
  \    A_2 =  \left( \begin{array}{cc}
                          b^2 & -ba \\ -ba & a^2
\end{array} \right)  \ ; \nonumber \\
 A_3 &=&   \left( \begin{array}{cc}
                          1 - (b^2/a^2) & 0 \\ 0 & 0
\end{array} \right)  \ .
\label{POVM-A2}
\end{eqnarray}
Such a POVM can never give a wrong result, and it gives an inconclusive
result when the outcome is $A_3$.
This POVM was found in \cite{Per93,EHPP} in the
context of distinguishing the
two states of~\cite{Ben92}).
It is the optimal process for obtaining a perfect conclusive outcome,
and a conclusive result is obtained
with probability $1 - (|a|^2 - |b|^2)$.
In our case, this is the probability of a successful teleportation.
Alice tells Bob whether she succeeded in teleporting the state by sending him
one bit, and in
addition to this bit,
Alice still has to send Bob the two bits for distinguishing the
four possible states (so he can perform the required rotation).
Alternatively, she can send him only one bit telling him whether he received
the state or not (as we explained for the case of fully entangled state)
loosing $\frac{3}{4}$ of the successful teleportations.

When used for distinguishing non-orthogonal states,
this POVM allows to get the optimal deterministic information
from two non-orthogonal states, although, on average,
it yields less mutual information
than the optimal projection measurement. In the same sense, on average,
the conclusive teleportation
does not yield the optimal average fidelity, but when it is successful --
the fidelity is one.

The conclusive teleportation process proves that any (pure) entangled state
presents quantum non-locality.
This fact can also be seen using
the filtering method~\cite{Procrust} when
applied to pure states.

\section{Arbitrary good conclusive bilocal teleportation
via mixed states}

In a perfect conclusive teleportation 
Alice performs a teleportation process which is sometime
successful, and when it is successful, the fidelity of the teleported state
is one. In an imperfect conclusive teleportation, Alice performs
a teleportation process which is sometime
successful, and when it is successful, the fidelity of the teleported state
is less than one but better than could be achieved with a standard
teleportation.

The original idea of teleportation
involves only one way classical communication
from Alice to Bob.  
We shall now extend 
\footnote{The most general teleportation channel
involving all local quantum operations
plus 2-way classical communication (LQCC) 
protocols 
was introduced in Ref. \cite{single}.}
it, 
allowing Bob to call Alice as well so that bilocal protocol is used.
Note that here we do  not consider the most general bilocal protocol
(the so called ping-pong protocol) but only allow Bob and Alice to operate
independently of the operation of the other. 
A ping-pong protocol could improve the probability of successful
projection (e.g., increase the $p'(p)$ described below),
by allowing several ``paths'' of successful 
distillation depending on the outcomes of
the measurements in each step of the protocol.
The communication (in our example) is just used to verify that the state was
teleported.
This generalization of teleportation makes sense,
as in many cases the classical communication is treated
as a free resource.

We have shown previously that a perfectly reliable conclusive
teleportation can be achieved when pure entangled states are shared.
We now show that it is possible to perform arbitrary good bilocal
conclusive teleportation when certain {\em mixed states} are used
(however, see the remark in the acknowledgements).
The arbitrarily good conclusive teleportation (which we call
``quasi-conclusive teleportation'') is not described by a particular POVM,
${\cal A} = \{A_1, \ldots A_m\}$, 
but by a series of POVMs 
${\cal A}^n = \{A_1(n), \ldots A_m(n)\}$, 
where $n$ is the index of this series.
For any $\epsilon$ we can find $n$ such that the POVM ${\cal A}^n$
yields fidelity better than $1-\epsilon$ for teleportation.
Yet, perfect fidelity cannot be achieved since the probability of success
goes also to zero when $\epsilon$ goes to zero.
Thus, we show that quasi-conclusive
teleportation is successfully done via mixed states!

We first purify the mixed state,
and then use it for teleportation.

Consider the state
\begin{equation}
\varrho_{p}=p|\Psi^{-} \rangle \langle \Psi^{-}|
+ (1-p) | 00 \rangle \langle 00 | \ , \ \ 0<p<1
\label{psi}
\end{equation}
which is a mixture of a singlet (with probability $p$) and a $|00\rangle$
state (with probability $1-p$).
Let the bilocal Alice and Bob action
be described in the following way:
\begin{equation}
\varrho_{p} \rightarrow \varrho' \equiv
\frac{V_1 \otimes W_1 (\varrho) V_1^{\dagger} \otimes W_1^{\dagger}
}{Tr(V_1 \otimes W_1 (\varrho) V_1^{\dagger} \otimes W_1^{\dagger})}
\label{gen} \ .
\end{equation}
It can be realized by performing generalized measurements
by Alice and Bob independently, i.e.,
Alice performs the measurement defined by the pair 
of operators $\{ V_1, V_2 \equiv \sqrt{I - V_1V_1^{\dagger}} \}$,
and Bob performs the measurement defined by the pair of operators
$\{ W_1, W_2 \equiv \sqrt{I - W_1W_1^{\dagger}} \}$.
[Alice's POVM
is the set 
${\cal A}=\{ A_{1}=V_{1}^{\dagger}V_{1}, A_{2}=V_{2}^{\dagger}V_{2} \}$,
and Bob's POVM
is the set 
${\cal B}=\{ B_{1}=W_{1}^{\dagger}W_{1}, B_{2}=W_{2}^{\dagger}W_{2} \}$.]
When the outcomes of both Alice and Bob is 1, 
which correspond to the first operator in
each lab ($V_1$ and $W_1$ respectively) the above transformation is
successfully done.

After getting the results of their measurements
Alice and Bob communicate via phone
to keep only those particles for which both results correspond
to the successful case. 
To show that quasi conclusive teleportation can be performed,
we define the sequence of POVM operators 
(in basis $\{ |0 \rangle |1 \rangle \}$:
\begin{equation}
          V_1(n)=  \left( \begin{array}{cc}
          (1/n) & 0 \\ 0 & 1  
          \end{array} \right)  \ ;
          W_1(n)=  \left( \begin{array}{cc}
          (1/n) & 0 \\ 0 & 1  
          \end{array} \right)  \ .
\label{AB}
\end{equation}
After the action of the corresponding
POVM the new state is  
$\varrho'=\varrho_{p'} \equiv p' |\Psi^{-} \rangle \langle \Psi^{-}|
+ (1-p') | 00 \rangle \langle 00 |$
with
the parameter $p'$ depending on the input parameter $p$ 
as follows
\begin{equation}
p'(p)=\frac{1}{1+\frac{1-p}{np}} \ .
\label{p'}
\end{equation}
The probability of successful transition
from $\varrho_{p}$ to $\varrho_{p'}$ is
\begin{equation}
P_{p \rightarrow p'}= \frac{1}{n^{2}}[1 + (n-1)p] \ .
\label{probability}
\end{equation}
Thus one can produce the state which has arbitrary good singlet fraction
(a fidelity with a singlet)
$F(\varrho_{p})=\langle \Psi^{-}_{12}| \varrho_{p}\Psi^{-}_{12} \rangle)$,
which obviously allows for arbitrary good conclusive teleportation.
The key point is
that the probability of successful
teleportation {\it decreases to zero} with fidelity of teleportation
(or equivalently singlet fraction) {\it going to unity}.
But it is nonzero for any required fidelity arbitrary close to perfect one.

One natural question is whether it is possible to make
teleportation arbitrary good via other mixed states.
In general the answer is negative.
In the case of Werner states (states in which a fully entangled state is
mixed with the completely mixed state), for instance, this
is a consequence of the fact that arbitrary good
conclusive distillation is impossible \cite{Popescu}.
In fact for those states the entanglement
fidelity cannot be increased.
Its  best value $F_{max}$ is the initial (before the
conclusive process) value $F_{0}$.
Thus, following~\cite{single},
the maximal teleportation fidelity
is equal to $\frac{2F_{0}+1}{3}$ and is
less than $1$ apart from the trivial case where the initial state is fully
entangled.

Another interesting question is 
whether it is possible to perform quasi-conclusive 
teleportation via mixed states 
with only one way classical communication.

This represents a more complicated
issue which requires a more complicated technical analysis,
and will be analyzed elsewhere.

\section{Summary}

In this paper we presented a new way of viewing the teleportation of an
unknown quantum state.  We showed that teleportation is 
a special and particular case 
of generating $\rho$-ensembles at a distance, hence, 
a special case of generalized EPR nonlocality (the HJW-EPR nonlocality). 
We believe that this view of
teleportation reduces some of the mystery of that process, and in
particular, explains why two classical bits can be sufficient for the
teleportation of a qubit. 
This work also showed the usefulness of the HJW
generalized EPR nonlocality, and their understanding that any
$\rho$-ensemble can be generated nonlocally.

We feel that understanding the connection between these two 
important forms of 
nonlocality improves much the understanding of entanglement.

Based on the connection between teleportation and generalized measurements, 
we presented the process of conclusive teleportation, 
a teleportation which is sometime
successful. We showed that any pure entangled 
state can be used to perform conclusive
teleportation with fidelity one, and more surprising, certain mixed states
can also be used to achieve conclusive teleportation with fidelity as close
to one as we like.

\section*{Acknowledgements}

The first sections (teleportation via POVMs and conclusive teleportation)
were presented before (but not published)~\cite{telePOVM}.
The idea of bilocal teleportation and the ability to perform
quasi-conclusive teleportation with mixed states were presented
in~\cite{single}, but the example we provide here is much 
simpler, and involves only two qubits. 
The connection between conclusive teleportation
and conclusive purification (or distillation) from a single pair
was obtained in~\cite{single} and the idea of conclusive distillation
appears earlier in previous works~\cite{Procrust}.

T.M.~would like to thank Asher Peres for helpful discussions.
Part of this work was done in the AQIP'99 conference.
The work of T.~M.~was supported in part 
by grant \#961360 from the Jet Propulsion Lab,
and grant \#530-1415-01 from the DARPA Ultra program.
The work of P.H. was supported by grant no. 2 PO3B 103 16 from the Polish 
Committee for Scientific Research.


\begin{thebibliography}{99}

\bibitem{BBCJPW} C. H. Bennett, G. Brassard, C. Cr\'epeau,
R. Jozsa, A. Peres and W. K. Wootters,
``Teleporting an unknown quantum state via dual classical 
and Einstein-Podolsky-Rosen channels''
{\em Phys. Rev. Lett.}, vol.~70, pp.~1895--1899, 1993.

\bibitem{Hels67} C.~W.~Helstrom, ``Detection theory and quantum
mechanics'', {\em Information and Control}, vol.~10, pp.~254--291, 1967.

\bibitem{Kholevo1} A. S. Kholevo, ``Information-theoretic aspects of quantum
measurement'', {\em Problems of Information Transmission},
vol.~9, pp.~110--118, 1973.

\bibitem{Kholevo2} A. S. Kholevo, ``Bounds for the quality of information
transmitted by a quantum communication channel'',
{\em Problems of Information Transmission},  vol.~9, pp.~177--183, 1973.

\bibitem{Helstrom} C. W.  Helstrom, {\it Quantum Detection and Estimation
Theory }, Academic Press, New York (1976).

\bibitem{Davies} E.~B.~Davies, "Information and quantum measurement'',
{\em IEEE Transaction on
Information Processing}, vol.~24, pp.~596--599, 1978.

\bibitem{Per93} A. Peres, {\it Quantum Theory: Concepts and
Methods\/}, Kluwer, Dordrecht (1993), Chapt.~9.

\bibitem{Ben-Div-95} C.~H.~Bennett and D.~DiVincenzo, 
{\it Quantum computing: Towards an engineering era?},
{\em Nature}, vol.~377, pp.~389--390, 1995.

\bibitem{Ben-Sho-98} C.~H.~Bennett and P.~W.~Shor, 
``Quantum information processing'',
{\em IEEE Transaction on
Information Processing}, vol.~44, pp.~2724--2742, 1998.

\bibitem{EPR}
A. Einstein, B. Podolsky and N. Rosen, ``Can quantum-mechanical description
of physical reality be considered complete?'', 
{\em Phys. Rev.} vol.~47, pp.~777--780, 1935.

\bibitem{Bell} J. S. Bell, 
{\em Physics}, vol.~1, pp.~195, 1964.

\bibitem{BMR} S. L. Braunstein, A. Mann and M. Revzen,
``Maximal violation of Bell inequalities for mixed states'',
{\em Phys. Rev. Lett.}, vol.~68, pp.~3259--3261, 1992.

\bibitem{deutsch85}
D. Deutsch, ``Quantum theory, the Church--Turing principle and the universal 
quantum computer,'' {\em Proc. Roy. Soc. London Ser. A.}, vol.~400, 
pp.~97--117, 1985.

\bibitem{ShorFactor} P. Shor, ``Polynomial--time algorithms for prime
factorization and discrete logarithms on a quantum computer,'' {\em SIAM
J. Computing}, vol.~26, pp.~1484--1509, 1997.

\bibitem{Grover} L. Grover, ``A fast quantum mechanical algorithm for
database search,'' in {\em Proceedings of 28th ACM Symposium on Theory
of Computing}, pp.~212--219, 1996. 

\bibitem{NatureL}
I. L.  Chuang, L. M. Vandersypen, X. Zhou,
D. W. Leung, S. Lloyd,
   ``Experimental realization of a quantum algorithm'',
{\em Nature},  vol.~393, pp.~143--146, 1998.

\bibitem{EPR-scheme} 
C.~H.~Bennett, G.~Brassard and N. D. Mermin,
``Quantum cryptography without Bell's theorem'',
{\em Phys. Rev. Lett.}, vol.~68, pp.~557--559, 1992; this work is
based on a previous scheme of
A. K. Ekert, ``Quantum cryptography based
on Bell's theorem'', {\em Phys. Rev. Lett.}, vol.~67,
pp.~661--663, 1991, which was the first to use entanglement for quantum key
distribution.

\bibitem{BHM96} E. Biham, B. Huttner and T. Mor,
``Quantum Cryptographic Network based on Quantum Memories'',
{\em Phys. Rev. A}, vol.~54, pp.~2651--2658, 1996.

\bibitem{ShorQEC} P. Shor,
``Scheme for reducing decoherence in quantum computer memory,''    
{\em Physical Review A}, vol.~52, pp.~2493--2496, 1995.
A.~M.~Steane, ``Error correcting codes in quantum theory'',
{\em Phys. Rev. Lett}, vol.~77, pp.~793--797, 1996.

\bibitem{ScienceCLSZ}
I.~L.~Chuang, R.~Laflamme, P.~W.~Shor, and W.~H.~Zurek,
``Quantum computers, factoring, and decoherence''
{\em Science}, vol.~270, pp.~1633--1635, 1995.

\bibitem{ShorFT} P. W. Shor, ``Fault-tolerant Quantum Computation,''
{\em Proc.~37th Annual Symposium on Foundations of Computer Science},
IEEE Computer Society Press, pp.~56--65, 1996.

\bibitem{ScienceZ}
E. Knill, R. Laflamme and  W. H. \.Zurek,
``Resilient quantum computation'',
{\em Science}, vol.~279, pp.~342--346, 1998, and references therein.

\bibitem{CRSS} A.~R.~Calderbank, E.~M.~Rains, P.~W.~Shor, and
N.~J.~A.~Sloane,
``Quantum error correction via codes over GF(4)''
IEEE Transactions on Information Theory {\bf 44}, pp.~1369--1387, 1998;
``Quantum error correction and orthogonal geometry'',
{\em Phys. Rev. Lett.}, vol.~78, pp.~405--408, 1997.

\bibitem{Rains} E.~M.~Rains, ``Quantum codes of minimum distance two'',
{\em IEEE Transactions on Information Theory}, vol.~45, pp.~266--271, 1999.

\bibitem{Caves} C.~M.~Caves, ``Quantum teleportation - A tale of two cities'',
{\em Science}, vol.~282, pp.~637--638, 1998.

\bibitem{ExTel}
D. Bouwmeester, J.-W. Pan, K. Mattle, M. Eibl, H. Weinfurter and
A. Zeilinger, ``Experimental quantum teleportation'',
{\em  Nature (London)}, vol.~390, pp.~575--579, 1997.
S.~L.~Braunstein, H.~J.~Kimble, 
 ``A posteriori teleportation'',
{\em Nature},  vol.~394, pp.~840--841, 1998.
D. Boschi, S. Brance, F. De Martini, L. Hardy and S. Popescu,
``Experimental Realization of Teleporting an Unknown Pure Quantum
State via Dual Classical and Einstein-Podolsky-Rosen Channels'',
{\em Phys. Rev. Lett.}, vol.~80, pp.~1121--1125, 1998.
A.~Furusawa, J.~L.~Sorensen, S.~L.~Braunstein, C.~A.~Fuchs, H.~J.~Kimble,
and E.~S.~Polzik,
``Unconditional quantum teleportation'',
{\em Science}, vol.~282, pp.~706--709, 1998.

\bibitem{HJW} L. P. Hughston, R. Jozsa, W. K. Wootters,
``A complete classification of quantum ensembles having a given density 
matrix'',{\em Phys. Lett. A}, vol.~183, pp.~14--18, 1993.

\bibitem{JP-DL} J. M. Jauch and C. Piron, 
``Generalized localizability'', 
{\em Helv. Phys. Acta.}, vol.~40,
pp. 559--570, 1967;
E. B. Davies and J. T. Lewis, 
``An operational approach to quantum probability'',
{\em Com. Math. Phys. },  vol.~17, 
pp.~239-260, 1970.

\bibitem{BB84} C.~H.~Bennett and G.~Brassard,  ``Quantum cryptography:
public key distribution and coin tossing'',
in {\em Proceedings of IEEE
International Conference on Computers, Systems and Signal Processing,
Bangalore, India} (IEEE, New York, 1984) pp.~175--179.

\bibitem{Ben92} C. H. Bennett,``Quantum cryptography
using any two nonorthogonal states'',
{\em Phys. Rev. Lett.}, vol.~68, pp.~3121--3124, 1992.

\bibitem{Gisin} N. Gisin, 
         ``Nonlocality criteria for quantum teleportation'',
{\em Phys. Lett. A.} vol.~210, pp.~157--159, 1996.

\bibitem{Fuc-Gra-99} C.~A.~Fuchs, J.~van de Graaf,  
``Cryptographic distinguishability measures for quantum-mechanical
states'',
{\em IEEE Trans. Inform. Theory},
vol.~45, pp.~1216--1227, 1999.

\bibitem{IvPe} I.~D.~Ivanovic, {\em Phys. Lett. A},
``How to differentiate between non-orthogonal states'',
vol.~123, pp.~257--259, 1987; A.~Peres,
{\em Phys. Lett. A}, ``How to differentiate between non-orthogonal
states'',vol.~128, pp.~19, 1988.

\bibitem{EHPP} A.K.~Ekert, B.~Huttner, G.M.~Palma and A.~Peres,
``Eavesdropping on quantum-cryptographical systems'',
{\em Phys. Rev. A.}, vol.~50, pp.~1047--1056, 1994.

\bibitem{Procrust}
N. Gisin,
``Hidden quantum nonlocality revealed by local filters'',
{\em Phys. Lett. A}, vol.~210, pp.~151--156,  1996.
C. H. Bennett, H. Bernstein, S. Popescu and B. Schumacher,
``Concentrating partial entanglement by local operations'',
{\em Phys. Rev. A}, vol. 53, pp. 2046--2052, 1996.

\bibitem{single}
M.~Horodecki, P.~Horodecki and R.~Horodecki, ``General teleportation
channel, singlet fraction and quasi-distillation'', quant-ph/9807091,
{\em Phys. Rev. A}, in press, 1999.

\bibitem{Popescu}
N. Linden, S. Massar and S. Popescu,``Purifying Noisy Entanglement Requires 
Collective Measurements'',
{\em Phys. Rev. Lett.}, vol.~81, pp.~2839--3279, 1998.

\bibitem{telePOVM} T. Mor, ``Tele-POVM: New faces of
teleportation'', quant-ph/9608005; presented
in a workshop, on Foundation of Quantum Theory, 
A Golden Jubilee event of the 
TIFR, September 1996, Bombay, India.

\end{thebibliography}
\end{document}